\documentclass[%
reprint,
superscriptaddress,
amsmath,amssymb,
aps,
pra,
]{revtex4-2}

\usepackage{placeins}
\usepackage{float}
\usepackage{caption}
\usepackage{graphicx}% Include figure files
\usepackage{dcolumn}% Align table columns on decimal point
\usepackage{bm}% bold math
\usepackage{booktabs}  % neatly formatting lines
\usepackage{placeins}
\usepackage{mathabx}
\usepackage{natbib}

\begin{document}
	%\bibliographystyle{unsrt}
	%\preprint{}
	
\title{An elementary method to determine the critical mass of a sphere of fissile material based on a separation of neutron transport and nuclear reaction processes}%

\author{S.K. Lamoreaux}%
\email{steve.lamoreaux@yale.edu}
\affiliation{%
Yale University, Department of Physics\\
New Haven, CT}%
	
\date{\today}% It is always \today, today,
	% but any date may be explicitly specified

\begin{abstract}
		
A simplified method to calculate the critical mass of a fissile-material sphere is presented. This is a purely pedagogical study intended to elucidate the historical evolution of criticality calculations and to help students interested in nonproliferation studies understand the basic physics of critical assemblies.

The method employs only elementary calculus and straightforward statistical arguments. The problem is formulated in terms of the threshold condition that the number of neutrons in the sphere does not change with time: the average neutron path length in the material must be long enough that fission produces sufficient neutrons to balance losses due to absorption and leakage through the surface.

This formulation separates the nuclear-reaction part of the problem from the geometry and mechanics of neutron transport. The only connection is the total path length, which, together with the distance between scatterings, determines the sphere radius. The approach yields an expression for the critical radius without the need to solve the diffusion equation.

Comparison with known critical masses shows agreement at the few-percent level. The analysis can also be applied to impure (including isotopically mixed) materials and can be extended to general neutronics estimates as a design guide or for order-of-magnitude checks of Monte Carlo N-Particle (MCNP) simulations. We compare the results with the Oppenheimer--Bethe criticality formula, with other calculations, and with a diffusion-equation approach using a new treatment of the boundary conditions.
		
\end{abstract}
	
\keywords{nuclear reactors, criticality calculations, neutron transport, neutronics, neutron multiplication}%Use showkeys class option if keyword
	%display desired
	
\maketitle
	
%\tableofcontents
\section{Introduction}
\newcommand{\Var}{\operatorname{Var}}
The analysis of criticality conditions presented here was developed as a curriculum component of a course in the Yale Department of Physics, ``Impact of the Atom,'' which focused on the Manhattan Project.
The method used to determine the critical mass of a fissile-material sphere employs only elementary calculus and straightforward statistical arguments. Due to the varied backgrounds of the students in the course, it was necessary to avoid advanced mathematical methods, such as solving the diffusion equation. Providing a fundamental understanding of what is needed to attain a critical assembly---without an undue mathematical burden---has proven useful, especially for students interested in nonproliferation studies and issues related to nuclear power.

This paper also includes a discussion of uranium enrichment. For example, enriching natural uranium to 20\% $^{235}$U is roughly halfway---in terms of separation stages---to weapons-grade material (approximately 90\% $^{235}$U). Beyond 20\%, further enrichment to weapons grade requires processing about 25 times less material. Thus, if it took one year to produce a metric ton of 20\% enriched material, then (in an idealized scenario) reconfiguring the serial cascade of separation stages into 25 identical parallel cascades (and assuming the earliest stages are already parallelized) could produce about 200 kg of weapons-grade material in roughly two weeks.

Pedagogical treatments for estimating the critical radius of a fissile-material sphere are available in a number of publications, for example, \cite{ajp1,ajp2,reed1}.
In the following, we develop a straightforward random-walk model to determine the critical radius. The model achieves surprisingly good accuracy and is applicable to isotopic mixtures or impure materials. The advantage of this method is that it leads to a compact closed-form expression for the critical radius without having to solve the diffusion equation.

Although the analysis is analytical, the use of a random-walk formalism is, in a sense, an application of Monte Carlo ideas: neutron transport is treated statistically rather than by solving a differential equation.

This approach can also be applied to back-of-the-envelope estimates for general neutronic experiments and can help build understanding of the voluminous information produced by large modeling programs, such as the Los Alamos MCNP software.

The goal of this paper is not to produce highly accurate critical-mass calculations that challenge well-known professional analyses, but rather to highlight the physical processes involved in a simplified and uncluttered fashion that is accessible to sophomore- or junior-level college students. Nevertheless, the results are astonishingly accurate.

Note that the criticality condition is simply the threshold at which a chain reaction can be maintained. To obtain efficient explosive energy, an assembly significantly greater than the threshold is required. Furthermore, the critical mass can be reduced by using so-called ``tampers,'' which back-reflect neutrons escaping from a sub-critical assembly. Although tampers are not discussed in this note, the methods introduced here can be readily modified to include them.
 
 \section{Overview of the Calculation Method}
 
There are three main simplifications in this approach. First, we assume that the essential reaction parameters---cross sections and the number of neutrons liberated per fission---are approximately energy-independent near 1 MeV, so that we may use values averaged over the fission-neutron energy spectrum. This was done in the Manhattan Project and leads to reasonably accurate results.

An instructive example is that the number of neutrons produced per fission decreases with neutron energy, whereas the fission cross section increases, so their product is comparatively constant (see Ref.~\cite{chadwick1}). The remaining approximations are that the neutron density is homogeneous within the material and that there is a position-independent root-mean-square distance that a neutron must travel to leave the sphere, as determined by the sphere radius together with scattering (diffusive) processes.

A fissile-material sphere is at the criticality threshold if the number of (free) neutrons $N$ in the sphere is constant in time. Here we specify $N$ without regard to how that state might be established experimentally.
Neutrons are continually absorbed, with some absorptions leading to fissions, each producing $\nu$ new neutrons while consuming the absorbed neutron that induced the fission.

Neutrons also escape through the surface of the sphere, since there is no containment mechanism. In the absence of other loss processes, the average time a neutron spends in the sphere is determined by the sphere radius together with scattering (diffusive) processes. In the absence of nuclear reactions, this defines a dwell time $\tau_w$ for any given neutron, implying a path length $\ell=\widebar v\tau_w$, where $\widebar v$ is the average velocity. The path length is thus a proxy for the time spent in the sphere. Without nuclear reactions, every neutron eventually leaves, with rate $\dot N=N/\tau_w$.

At the criticality threshold, the required path length $\ell$ is determined purely by the nuclear-reaction cross sections. Note that $\ell$ need not be a straight-line distance and, in general, is not set by scattering.

There are thus three components to the overall analysis: (i) fission and non-fission capture reactions that can occur as a neutron travels a distance $\ell$; (ii) neutron loss through the sphere surface; and (iii) the relationship between $\ell$ and the sphere radius as determined by a random walk due to neutron scattering.

Because loss through the surface is impeded by scattering within the material, a higher scattering probability tends to keep neutrons in the material longer, increasing the likelihood of fission and thereby reducing the critical size. Equivalently, the required $\ell$ can be achieved with a smaller-radius sphere when diffusive processes impede neutron motion.
 
\section{Derivation of the critical radius of a sphere}

\subsection{Determination of $\ell$ from nuclear reaction cross sections}

We first consider only neutron reactions in which a neutron is absorbed and lost; however, some absorptions lead to fission and the production of more neutrons.  Scattering, as a mechanical process associated with neutron motion or transport, will be discussed later.

We consider a gas of $N$ neutrons within a sphere of material of nuclear number density $n$.
The neutrons can suffer either non-fission absorption
with cross-section $\sigma_a$ or cause a fission (cross section $\sigma_f$) with $\nu$ neutrons
being released per fission. Neutrons constantly leak out of the sphere at the end of their paths if not absorbed during travel in the sphere.

The definition of the reaction cross section is that if a neutron
travels a distance $\ell$ through the material, the probability of traversing this distance without suffering either absorption or causing a fission is
\begin{equation}
	P(\ell)=e^{-n(\sigma_a + \sigma_f)\ell}.
\end{equation}

Recall that the distance $\ell$ will in general not be a straight line because a neutron may
suffer several nearly elastic scatterings before being absorbed or causing a fission. These
scatterings have no direct bearing on the fission reaction. The consideration of scattering is necessary to convert $\ell$ into the radius of the sphere required for criticality, as will be determined in the next section.

It is convenient to define 
\begin{equation}
\sigma_a+\sigma_f=\sigma_0
\end{equation}
as the sum of all absorption processes.

Now, the probability $P_0$ that an absorptive reaction {\it did} occur within the path length $\ell$ is given by
$P_0=1-P(\ell)$. 
Of all the reactions that occur, the fraction that
will cause fissions is simply $\sigma_f/\sigma_0=P_f/P_0$,
where, as in the introduction, the probability of a fission reaction over the distance $\ell$ is $P_f$, and the number of fission neutrons produced, $N_f=N\nu P_f$:

Let us consider the probabilities of what can happen {\it on average} when there are $N$ neutrons in the core, in one period $\tau_w$, which corresponds to a neutron traveling a distance $\ell$:
\begin{enumerate}
	\item Fissions will occur, producing a number of neutrons   
\begin{equation}
N_f=\nu N\left(\frac{\sigma_f}{\sigma_0}\right)\left(1-e^{-n\sigma_0\ell}\right).\nonumber
\end{equation}
\item Neutrons will be absorbed, losing a number
\begin{equation}
N_0=NP_0=N\left(1-e^{-n\sigma_0\ell}\right)\nonumber.
\end{equation}
\item $N$ neutrons will leave the sphere, according to the definition of $\tau_w$ and $\ell$.
\end{enumerate}

The threshold of criticality is defined by $N$ being constant. Thus, we must have an ensemble average such that with each period $\tau_w$, the
number of neutrons produced by fission equals the loss by emission (neutrons exiting through the sphere surface) and by absorption,
\begin{equation}\label{equilibrium}
N_f - N_0 - N =0.
\end{equation}
The neutron emission rate $N/\tau_w$ from the sphere is reflected by $-N$ in the above equation.

It might seem strange that $N$ is a constant and that the emission, or ``leakage'' rate, remains constant in the presence of other processes. This is simply as expected with $N$ specified. From kinetic theory, the current $I$ through the surface is $I= N\widebar v A/4V$, where $\widebar v$ is the average velocity, $A$ is the surface area, and $V$ is the volume, and by our definitions $N=I\tau_w$ as $\dot N/N=1/\tau_w$. The presence of other processes does not affect this ``leakage'' rate. Alternatively, the entire problem could be cast in terms of rate equations (see Appendix F). There is no pressing need to do that.

Dividing Eq. (\ref{equilibrium}) through by $N$, which we can do because the system is linear, we arrive at the condition for threshold criticality as
\begin{equation}\label{critcond}
\left[\nu\frac{\sigma_f}{\sigma_0}-1\right]\left[1-e^{-n\sigma_0\ell}\right]=1.
\end{equation}
This relationship determines the critical distance $\ell_c$ for which the number of neutrons in the sphere remains constant,
per Eq. (\ref{equilibrium}).
From this relationship, it is straightforward to solve for $\ell_c$ as
\begin{equation}
 \label{ellc}
 \ell_c=-\frac{1}{n\sigma_0}~\ln\left[1-\frac{\sigma_0}{\nu~\sigma_f -\sigma_0}\right].
 \end{equation}
 
The critical size of the system is determined by the condition that, {\it on average}, the total number of neutrons remains constant. Within the approximations adopted here, there is no need to average over the full distribution of possible neutron path lengths.

Note that if the condition $\nu\sigma_f/\sigma_0>2$ is not met, a critical assembly is not possible.

 At this point, the nuclear reaction part of the problem is complete, so next we need to consider how the mechanics of neutron random walks can be used to relate $\ell_c$ to the critical radius $R_c$.
 
\subsection{Neutron random walk}

 As a neutron travels through the fissile material, it will follow a random walk with an average step size 
  \begin{equation}\label{ells}
\ell_s=\frac{1}{n\sigma_{tr}},
 \end{equation}
 where $\sigma_{tr}$ is the transport scattering cross section. This will be elaborated later in this
 section.
 Now, if the displacement of the neutron from its starting point exceeds some geometric dimension $\ell_0$, then the neutron will be lost from the system. This characteristic dimension is developed in the following subsection but could, 
 for example, be the average distance from any point inside the sphere to its surface. Irrespective of how this is formulated, the theory of random walks indicates that the number of steps $N_s$ required to attain $\ell_0$ is 
 determined by
 \begin{equation}
 \label{Ns}
\ell_0^2~\propto~ N_s\ell_s^2.
\end{equation}

In Appendix A, it is shown that the average squared distance from any point within a sphere of radius $R$
to the surface is $4R^2/5$. As shown in Appendix B, there is a modest correction factor because the square root of the average squared displacement from a random walk overestimates the actual average displacement.  
The factor of 4/5 and this correction factor can be combined together into a dimensionless scale factor $\epsilon$, 
so we write $\ell_0^2 = R^2/\epsilon^2$ and cast Eq. (\ref{Ns}) as 
\begin{equation}
\label{wd}
R^2 = N_s \epsilon^2 \ell_s^2.
\end{equation} 
It is shown in Appendix B that $\epsilon=\sqrt{10/4\pi}=0.89$. Given that its derivation presents a new technical complication, one could skip the derivation and introduce $\epsilon$ as a calculable correction factor, or simply omit this factor with the resulting critical masses about 30\% too large.    

\subsection{Neutron travel distance to escape and the critical radius} 

During the random walk of $N_s$ steps described above, the neutron will travel a linear distance $\ell=N_s\ell_s$.
If $\ell=\ell_c$, then the sphere will be just critical, that is, the criticality is obtained when

\begin{equation}
N_s\ell_s =\ell_c.
\end{equation}

This determines $N_s=\ell_c/\ell_s$, which can be used in Eq. (\ref{wd}) to give

\begin{equation}
R_c^2=\epsilon^2 \ell_c\ell_s.
\end{equation}

Combining this with Eq. (\ref{ellc}) gives the critical radius $R_c$ 
as
\begin{equation}\label{rc}
R_c={\epsilon \sqrt{\ell_s\ell_c}}=\frac{{\epsilon}}{n\sqrt{\sigma_{tr}\sigma_0}}\left[-\ln\left(1-\frac{\sigma_0}{\sigma_f\nu-\sigma_0}\right)\right]^{1/2}.
\end{equation}
The critical mass of the sphere is then $m_c=4\pi\rho R_c^3/3$, where $\rho$ is the usual mass density.

(In Appendix D the foregoing logic is applied to the diffusion equation, leading to a formula nearly identical to Eq. (\ref{rc}). The comparison is simplified with a new treatment of the boundary condition that leads to a quadratic equation.)

This is the essential result of our analysis. The transport scattering cross section $\sigma_{tr}$ is considered further below.
The principal parameters and variables used in this analysis are listed in Table 1 and discussed in the next Section.

To summarize the results so far, there are two distinct aspects of the
criticality problem. The first is to determine the path length that the neutron travels within the sphere, which must be long
enough for fission-induced neutron production to be great enough to replace those lost through absorption and emission by the sphere. 
The second is the diffusive process that impedes neutrons from leaving the sphere and determines the radius
required to satisfy the first criterion.  A sphere slightly larger than this is a critical mass, or, 
if a sphere that is slightly smaller is sufficiently compressed, it will become critical because the
critical radius is inversely proportional to the density. A discussion of compressed cores can be found in Chapter 2 of Ref. \cite{reed1}.

\begin{table*}[t!]
\caption{Principal parameters and variables.}

\begin{center}
\resizebox{.5\linewidth}{!}{
%\begin{table*}[h!]
%\caption{Principal parameters and variables.}

%\begin{center}
%\resizebox{.8\linewidth}{!}{
\begin{tabular}{cc} \toprule
	{Symbol}&{Definition} \\ \midrule
	$n$& Nuclear number density \\
	$\rho$&Mass density\\
	$m_n$&Neutron mass\\
	$A$&Atomic weight; mass number\\
 $\epsilon$&Combined random walk and geometry correction factor\\
	$P$&Probability that a neutron does not undergo a reaction \\
	$\nu$& Number of neutrons per fission \\
	$N_s$& Number of times a neutron scatters \\
	$N$& Number of neutrons in the sphere \\
    & \ \ (arbitrary constant at equilibrium) \\
	$\sigma_f$& Fission cross section\\
	$\sigma_a$& Absorption (without producing fission) cross section \\
    $\sigma_0$ & Total absorption cross section $\sigma_a+\sigma_f$\\
    $\sigma_e$& Elastic scattering cross section\\
    $\sigma_i$ & Inelastic scattering cross section\\
	$\sigma_{tr}$&Transport scattering cross section $\sigma_e+\sigma_i/2$ \\ 
	$\ell$&Distance of neutron travel through material\\
	$\ell_c$& Distance of neutron travel for criticality \\
	$\ell_s$& Mean distance between scatterings\\
 $\ell_0$& Mean distance from point inside sphere or mass to surface\\
	$R$&Sphere Radius\\
	$R_c$& Critical radius\\
	$m_c$&Critical mass\\
	\bottomrule
\end{tabular}}
\end{center}
\end{table*}

It remains to specify the transport scattering cross section $\sigma_{tr}$ introduced in relation to Eq. (\ref{ells}). Elastic scattering at MeV energies is typically forward-peaked; small-angle deflections contribute little to diffusive transport, so the usual angle-averaged elastic cross section is not, by itself, the appropriate quantity (inelastic scattering is approximately isotropic).

In elementary diffusion theory, forward peaking is accounted for by weighting the differential cross section with an additional factor $1-\cos\theta$ in the angular average.\cite{diffusion} Since ENDF-style tabulations often do not list this transport-weighted quantity explicitly, we introduce an effective $\sigma_{tr}$ suitable for back-of-the-envelope estimates.

Appendix C provides a heuristic estimate showing that, for the present purposes, the transport scattering cross section in Eq. (\ref{rc}) can be approximated by
 \begin{equation}
\sigma_{tr}\approx \sigma_e+\frac{1}{2}\sigma_i,
\end{equation}
where $\sigma_e$ and $\sigma_i$ are, respectively, the angle-averaged cross sections for elastic and inelastic scattering. 

In many cases, the cross sections related to fission and neutron absorption
do not vary much with energy in the 100 keV to 1 MeV range, which means that we can
use single representative values for them for computational purposes. Exceptions to this are $^{232}$Th and $^{238}$U, 
which have fission thresholds of about 1-2 MeV. Neutrons released in fissions generally have average kinetic energies of around 1-2 MeV, so even slight moderation with scattering can have a large effect for these isotopes, as will be discussed.

As an aside, it is worth pointing out some realistic complications that have been intentionally omitted. 

Because this note is intended to illustrate the fundamental physics by the simplest means possible, many details that are important to a realistic system are not included. This calculation applies to a bare sphere, which is the simplest critical system.

For example, including a neutron-reflecting "tamper" can significantly reduce the critical mass. In the Fat Man bomb, $^{239}$Pu together with a natural uranium tamper reduced the critical mass to 6.2 kg. Although this was the core mass, the system was made subcritical through the use of a lower-density phase of plutonium, resulting in an overall subcritical neutron multiplication factor of 0.78. Implosive compression increased the density by a factor of 2.5, which, together with the tamper, produced a system with a neutron multiplication factor approaching 4.

\section{Applications}

The parameter values for $^{235}\textmd{U}$, $^{238}\textmd{U}$, and $^{239}\textmd{Pu}$ are listed in Table II.
The values for $^{235}\textmd{U}$ correspond to a neutron energy of 1.5 MeV and are adopted from ENDF/B-VIII.0 files, as reported in Ref.~\cite{chadwick1}, Table~1, p.~S32.

The average neutron energy for a $^{239}$Pu near-divergent chain reaction in a spherical core is about 1.9 MeV, as reported in Ref.~\cite{pue}. ENDF/B-VII.0 values at this energy are given in Table II.

The parameters for $^{238}\textmd{U}$ require special treatment, as discussed below, because the (n,f) cross section has a threshold near the peak of the prompt-fission spectrum. The values listed are based on refs. \cite{U2382,238U} and are intended as representative averages over a $^{235}$U prompt-fission spectrum, with an additional effective reduction of $\sigma_f$ to account (crudely) for modest down-scattering in large, low-enrichment systems.

Calculated critical radii and masses from Eq. (\ref{rc}) appear in the next to last two lines, 
with critical masses from a full MCNP6 calculation (ref. \cite{chadwick1}, p. S32, and \cite{pucross}), using ENDF/B-VIII.0 cross sections given in the last line.

\begin{table*}[t!]
\caption{Parameter values. See text for comments regarding $^{238}\textmd{U}$; in particular, the fission cross section compared to the fission-spectrum average in \cite{U2382} is reduced by a factor of approximately 0.8 due to moderation effects. 1 barn = $10^{-24}{\rm cm}^2$}
\begin{center}

\resizebox{.4\linewidth}{!}{
\begin{tabular}{ccccc} \toprule
	{Parameter}&{$^{235}\textmd{U}$} &{$^{238}\textmd{U}$} & {$^{239}\textmd{Pu}$} & Units\\ \midrule
	$\nu$ & 2.57 & 2.47 & 3.156 &  \\
	$\sigma_e$ & 3.557 & 4.827 & 3.33 & barns \\
	$\sigma_i$ & 1.926 & 2.541 & 1.85 & " \\
    $\sigma_{tr}=\sigma_e+\frac{1}{2}\sigma_i$&4.52&6.10&4.25 & "\\
	$\sigma_a$ & 0.082 & 0.12 & 0.0172 &" \\
	$\sigma_f$ & 1.24 & 0.24$^{(*)}$ & 1.96 & " \\
    $\sigma_0=\sigma_a+\sigma_f$ &1.322&0.36&1.98 & "\\
	$\rho$ & 18.7 & 18.9 & 19.61 & ${\rm g} \, {\rm cm}^{-3}$ \\
	$R_c$ & 8.48 & --- & 4.96 & cm \\
	$m_c$  (Eq. 11)& 47.8 & --- & 10.00 & kg \\
        $m_c$  (MCNP6) &46.4$\pm$ 1.7&---&10.2 & kg\\
	\bottomrule
\end{tabular}}
\end{center}
\end{table*}

In the case of $^{239}$Pu, the critical mass is 10.0 kg, calculated using Eq. (\ref{rc}) with ENDF/B-VIII.0 parameters at
1.9 MeV (footnote h, p. S32, ref. \cite{chadwick1}), in excellent agreement with the MCNP6 value of 10.2 kg. For parameters at 1.55 MeV, the critical mass increases to 11.7 kg, illustrating the dependence of the critical mass on the average neutron energy. The mass density adopted depends on the phase of plutonium involved; the so-called high-density alpha phase is assumed here.

\subsection{Extension to imperfect enrichment}

For $^{235}$U, the calculated critical mass of $\sim$ 47.8 kg is somewhat less than the 49 kg reported in the original
paper cited in \cite{cm}, and is slightly greater than 46.4 kg cited in \cite{chadwick1}. Regarding the 49 kg value, the critical mass given in the reference refers to an enrichment of 93.71\% $^{235}$U. This is typical of the maximum enrichment used in weapons because further
enrichment by gaseous diffusion or centrifuges also increases the concentration of $^{234}$U, which has a relatively high absorption cross section; its absorption is significant enough that it cannot be ignored in a critical-mass calculation. For now, we have ignored the effect of $^{234}$U, which has
a very low natural abundance; the effect of $^{238}$U can be included by weighting the parameters by their relative concentrations.

The effect of neutron moderation has not been considered important so far. However, in the case of a large mass of material with a high concentration of $^{238}$U, a neutron will suffer many collisions before it is lost through the surface or absorbed. Although
the average energy change per scattering is small, it is important for $^{238}$U because of the low fission energy threshold.  The energy change per scattering on average is\cite{echange} 
\begin{equation}
1-\frac{\overline{\Delta E}}{E_0} \approx \frac{2A}{(A+1)^2}=8\times 10^{-3}.
\end{equation}
Taking a sphere of low-enrichment uranium (in $^{235}$U), as we will show, the critical mass is around 800 kg, giving a radius of 20 cm. The scattering length for a neutron in $^{238}$U is 3.5 cm, and in a random-walk displacement of 20 cm a neutron will scatter at most about 30 times if it starts from the center and escapes through the surface. Given that the rate of surface loss is roughly equal to the rate of absorption and induced fission, on average there will be about 15 scatterings, leading to a reduction in average energy by a factor of about 0.86.

The threshold incident neutron energy for $^{238}$U fission is approximately 1.5 MeV; below this energy, the cross section falls rapidly to zero, while above it attains a value of about 0.5 barn and is approximately constant up to 7 MeV. \cite{russ} It is straightforward to numerically integrate the overlap of the cross section by assuming a step function for the energy dependence.
The Watts spectrum for $^{235}$U fission neutrons is
\begin{equation}
 n(E)=C e^{-E/a}\sinh\!\left(\sqrt{bE}\right),
\end{equation}
where $E$ is the fission-neutron energy (MeV), $C$ is a normalization constant, $a=0.988$ MeV, and $b=2.249$ MeV$^{-1}$. As a rough correction for the reduced average neutron energy, we adjust $a$ and $b$ accordingly and numerically integrate the spectrum to estimate the fraction of neutrons above the $^{238}$U fission threshold (1.5 MeV). Integrating from 1.5 to 100 MeV gives a fraction of 0.488, so the effective fission cross section is $(0.488)(0.5~\text{b})\approx0.24$ b. This is lower than the $\sim0.30$ b prompt-spectrum-averaged value reported in \cite{U2382}. While crude, including this reduction is important; otherwise, our enrichment-dependent critical-mass estimates deviate substantially from more accurate calculations. Identifying this mechanism is a useful pedagogical point.

\begin{table*}[t!]
\caption{The effect of enrichment on the $^{235}$U critical mass, with a $^{238}$U fission cross section of 0.24 b and an average of 2.281 prompt neutrons per $^{238}$U fission. NV indicates that no value was listed in \cite{Glaser}, while x indicates that a critical mass is not possible.}

\begin{center}
\resizebox{.4\linewidth}{!}{

\begin{tabular}{ccc} \toprule
	\ \ Enrichment \% \ \ \ \ & $m_c$ kg &$m_c$ kg \cite{Glaser} \\ \midrule
100&47.8& NV\\
93&51.5&53\\
70&71&87\\
45&129&185\\
30&256&367\\
19.75&916&782\\
19&2300& x\\
18.9&x&x\\
\bottomrule
\end{tabular}}
\end{center}
\end{table*}

At an enrichment of 19.75\%, with $\sigma_f=0.24$ and $\nu=2.15$, our analysis produces a critical mass of 916 kg, which is in reasonable agreement with 800 kg cited in Ref.~\cite{cm}.
However, the literature reports widely varying values for lower enrichment levels. In a 2006 paper, Glaser
examined the proliferation potential of uranium enriched to various levels using the then-current MCNP4C code,
estimating a critical mass of 782 kg for 19.75\%.\cite{Glaser}

\begin{figure}[h]
	\includegraphics[scale=0.6]{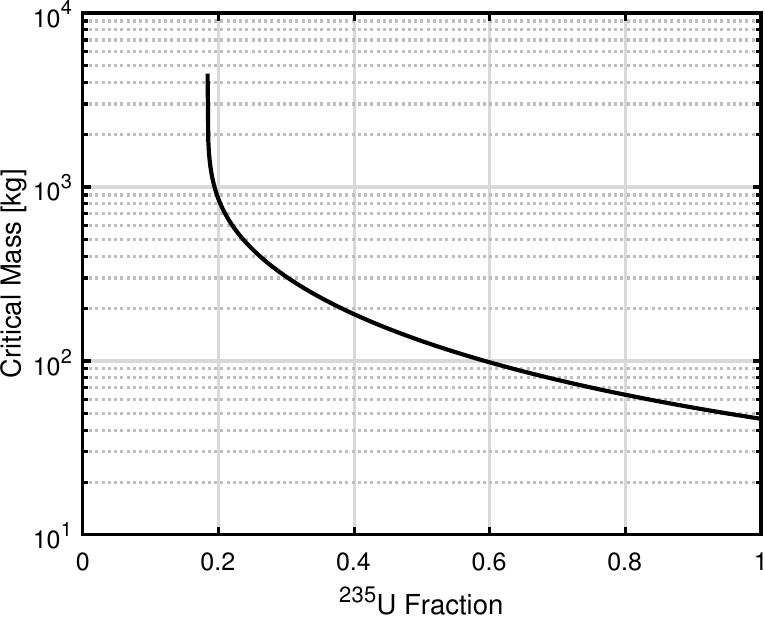}
	\vspace*{+5mm}
 \caption{Critical mass for $^{235}$U as a function of isotopic purity, with $^{238}$U the only contaminant, $\bar \nu=2.281$ and $\sigma_f=0.25$ b.}
\end{figure}
 \subsection{$^{234}$U as a contaminant}
 
As noted above, the presence of $^{234}$U can significantly increase the critical mass. The n-$\gamma$ absorption cross section of $^{234}$U is relatively large, $\sigma_a\approx0.2$ b, compared to that of $^{235}$U, which is 0.082. However, the fission cross section (accounting for the $^{234}$U fission neutron energy threshold of 0.9 MeV), neutrons per fission, and transport cross section are similar to those of $^{235}$U. The effect of $^{234}$U on the critical mass can be determined as a function of the relative concentrations of the $^{234}$U, $^{235}$U, and $^{238}$U isotopes, which depend on the number of enrichment stages.

\begin{figure}[h]
	\includegraphics[scale=0.6]{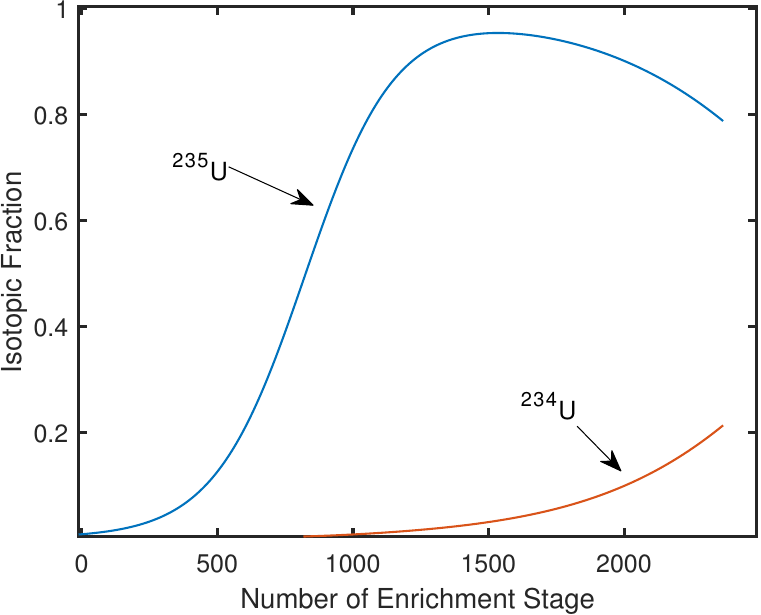}
	\vspace*{+5mm}
 \caption{Isotopic fraction of 234 and 235 as a function of the number of enrichment stages.\label{isofrac}}
\end{figure}

\begin{figure}[h]
	\includegraphics[scale=0.6]{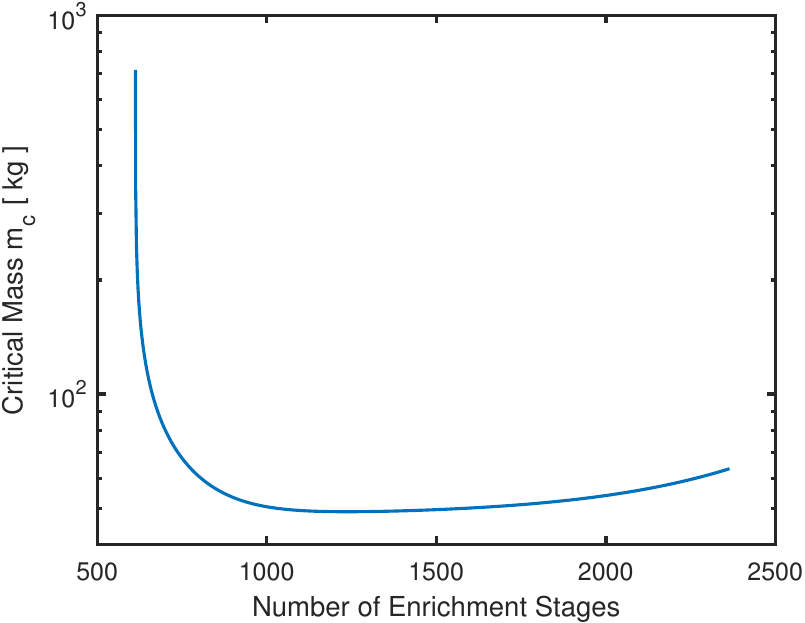}
	\vspace*{+5mm}
 \caption{Critical mass as a function of number of enrichment stages, starting with natural U (the atomic density is assumed independent of $A$). The critical mass at the minimum is about 50 kg.\label{isocrit} }
\end{figure}

Assuming ideal diffusion barriers, the transport rate through a barrier scales as $1/\sqrt{A}$, where $A$ is the atomic mass, and multiple barriers can be thought of as a series of flow impedances that depend on $A$. The total rate, normalized to unity for $^{238}$U, for $N$ barriers in series is
\begin{equation}
R=1 + C_{0,234}\left[\sqrt{238/234}\right]^N+C_{0,235}\left[\sqrt{238/235}\right]^N.
\end{equation}
The isotopic concentrations that are collected are
\begin{eqnarray*}
C_{N,234}&=&\frac{C_{0,234}[\sqrt{238/234}]^N}{R}\\ C_{N,235}&=&\frac{C_{0,235}[\sqrt{238/235}]^N}{R}.
\end{eqnarray*}
The isotopic fractions as a function of $N$ are shown in Fig.~\ref{isofrac}, and the effect on the critical mass is shown in Fig.~\ref{isocrit}, where the starting concentrations for 234 and 235 are $5.5\times 10^{-6}$ and $7.2\times 10^{-3}$, respectively. There is a broad minimum at $N\approx 1250$, after which the higher n-$\gamma$ absorption cross section of $^{234}$U causes the critical mass to increase. Compared to the case without $^{234}$U, the critical mass is slightly higher (about 50 kg) and similar to the 49 kg reported in the original paper cited in \cite{cm}.
It should be noted that the parameters for $^{234}$U are not as well studied as for $^{235}$U and $^{238}$U, so the results here are approximate.

\section{Number of first-generation neutrons produced as a function of sphere radius}

In this section, we consider the number of first-generation neutrons that are produced and that, on average, reach the mean distance $\ell_0$
to the edge of the sphere after a single neutron is placed randomly within the sphere.

The path length of travel in the material is determined approximately (without the $\epsilon$ correction factor) by:
\begin{equation}
\ell=N_s\ell_s=\ell_s(R^2/\ell_s^2)=R^2/\ell_s=n\sigma_{tr}R^2,
\end{equation}
where $\ell_s$ is again the mean free path for scattering. The number of neutrons  $N$ produced per neutron  is then
\begin{equation}
N=\left(\frac{\sigma_f}{\sigma_0}~\nu-1\right)\left[1-e^{-n^2\sigma_0\sigma_{tr} R^2}\right].
\end{equation}
Figure 2 shows $N$ as a function of $R$ for $^{239}$Pu.  With scattering in play, $N = 1$ at the critical radius
computed in the previous section, 5.2 cm. For a core of two critical radii, virtually every possible secondary neutron
that can be produced is produced.  

The effect of turning off scattering is also shown, so $\ell\approx R$
and the exponent becomes $n\sigma_0 R$. This indicates that scattering for the case of $^{239}$Pu decreases the free-travel radius by a factor of about 5, or a factor of 125 reduction in critical mass.

\begin{figure}[h]
	\includegraphics[scale=0.6]{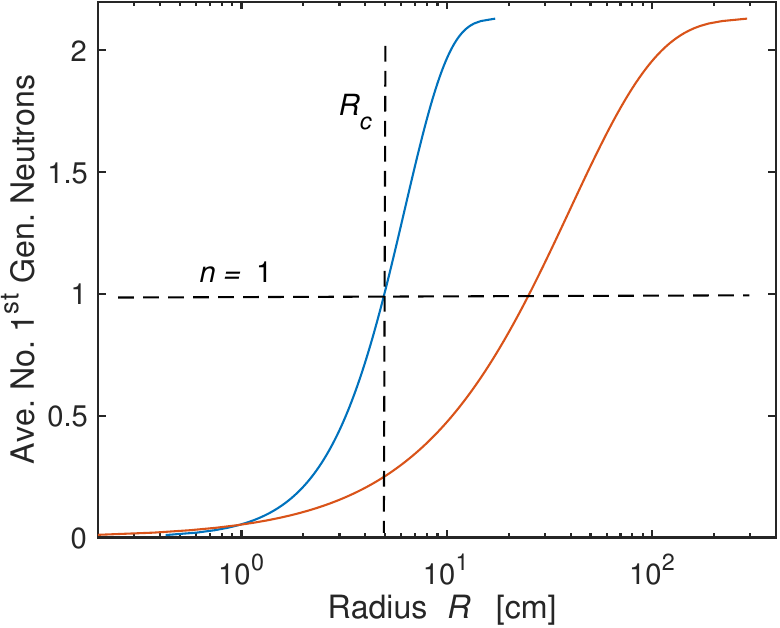}
\caption{Number of first-generation neutrons produced as a function of sphere radius, with (blue; left curve) and without (red; right curve)  the effects of scattering included, both without including neutrons from secondary fission, for $^{239}$Pu.}
\end{figure}

\section{Neutron production probability}

When a sphere is at the criticality threshold, introducing $N$ neutrons into the sphere will result (on average) in a steady total neutron-emission rate of $N/\tau_w$.

In the derivation of the mean criticality condition, $N$ was treated as a specified (equilibrium) average. Here we instead consider fluctuations about that equilibrium by following the stochastic neutron production from a single excess neutron.

The variance in the number of neutrons produced per time interval $\tau_w$ is of order the square root of the expected number. For the single-neutron case considered above, the expected number is $N=1$.
The probability of generating $k$ neutrons is given by the Poisson distribution,
\begin{equation}
P(k,N)=\frac{N^{k}e^{-N}}{k\ !}.
\end{equation}
For $P(0,1)=P(1,1)=e^{-1} = 0.368$, and the probability of generating two
or more neutrons is $P(1, k\geq 2)=1-2e^{-1}=0.26$.

What is more interesting is the probable {\it number} below $N$ and the number above $N$, $N_b$ and $N_a$ respectively, and the number at $N$, $N_N$,
\begin{eqnarray*}
N_b&=&\sum_{k=1}^{N-1}kP(k,N);\\  N_a&=&\sum_{k=N+1}^\infty kP(k,N);\\ 
N_N&=&N~P(N,N).
\end{eqnarray*}
For $N=1$, $N_b= 0$ while $N_a=0.632$ and $N_1=0.368$ (the sum is equal to $N=1$). However, on average $N_a$ is larger than $N_b$, so fractionally it appears that there is a tendency for more neutrons to leak out than remain.

That the Poisson distribution is skewed is well known; in the large-$N$ limit, the difference between the totals above and below $N$ is proportional to $\sqrt{N}$. 

For a larger number, say $N=10$, $N_b=3.33$, $N_a=5.42$, and $N_N=1.25$, which sum to $N=10$. As production in the sphere and leakage from the sphere are separate random processes that are coupled only through $N$, there are independent chances of upward and downward fluctuations for both production and loss. This implies that $N$ itself must fluctuate, because the change in $N$ is the difference of two random processes with the same mean. In the case of $N=10$, the difference $(N_a-N_b)/2\approx 1$, and if we consider $\Delta N$ between successive intervals of $\tau_w$, there is a variance in $N$---for both the production and loss processes---of about 1. Therefore, $N$ undergoes a random walk, and if it reaches zero, the process terminates. The random-walk step size is roughly $\sqrt{N}$, so as $N$ changes, so does the characteristic scale of its fluctuations.

The Poisson character of the neutron number distribution, even in the presence of multiple ignored random processes (path length, starting position, neutrons per fission, etc.), is expected based on the Central Limit Theorem.\cite{clt}

In the case of a subcritical sphere, i.e., when the right-hand side of Eq.~(\ref{critcond}) is less than one by a small parameter $\eta$, the number $N$ will decay over time on a longer timescale $\tau\approx\tau_w/\eta$. If $N_i$ neutrons are introduced into the sphere, the total number emitted is
\begin{equation}
N_e=\frac{N_i}{\tau_w}\int_0^\infty e^{-\eta t/\tau_w }dt=\frac{N_i}{\eta}.
\end{equation}
Studies of this effect were pioneered by Feynman et al. during the Manhattan Project \cite{galison}, and further developed by Prasad and Snyderman (see, e.g., \cite{snyderman}). The fact that multiplication measurements can provide a ``nondestructive'' determination of a system's approach to criticality (i.e., $\eta=0$) helps explain the appeal of these dangerous measurements. In addition to the total neutron production, fluctuations during the process---as in the threshold case---also provide information about the system. It is easy to imagine that if $0<\eta \ll 1$, relatively intense random neutron bursts would occur.

The analysis presented here provides simplifications toward the understanding and calculation of such fluctuations and is worthy of further study in the context of determining the critical mass.

\section{Fission Time Evolution}

Here we make an order-of-magnitude estimate of the time 
to fission one mole of $^{239}$Pu nuclei with negligible loss of neutrons through the surface.
This would give a yield of about 4 kilotons; Figure 2 shows that a core of two critical radii
satisfies the no-loss condition.  
In this case, the net number of fission neutrons produced per generation is, accounting for the one neutron that is needed to induce each fission and assuming that non-fission absorption is small (well satisfied for $^{239}$Pu),
\[(3.156-1)\times\sigma_f/\sigma_0=2.13.\]
The number of generations $N_g$ to cause a mole of fissions when starting from a single neutron is then $2.13^{N_g}=6.02\times 10^{23}$,
or
$N_g\sim 72$.  The mean distance traveled between fissions is
$1/n\sigma_f=11\ {\rm cm}$.
(Scattering occurs over a smaller distance, but this does not matter so far as computing a timescale is concerned.)
The total distance traveled is consequently $72\times 11\sim 800$ cm for the uncompressed metal.  The velocity of a 1 MeV neutron is $1.4\times 10^9$ cm/s, so the time scale is $800/1.4\times 10^9=0.57\times 10^{-6}$ s or 0.57 $\mu$s. 
Nuclear explosions are microsecond-scale events. 

These numbers give a sense of the required timing accuracy of an implosion system.  For $^{235}$U, the time will be a factor of two longer, but is still on the order of a microsecond. However, because of the low neutron background in $^{235}$U (compared to $^{239}$Pu as produced in a reactor and which is
generally contaminated with spontaneously fissile $^{240}$Pu), the components can be assembled in a more leisurely way.

The fission of one mole of $^{239}$Pu releases an energy equivalent of about 4 kilotons of TNT.  
The energy released in the Trinity test was about 20 kilotons, or about five moles worth of fissions. 
The core mass was 6.3 kg ($m_c$ was lowered by using a reflector, also known as a tamper), corresponding to 26 moles, 
implying an efficiency of 4/26 or 15\%. The implication is that the supercritical core had disassembled a few
generations after the first mole of fissions.  Calculating the full dynamics of the fissioning core is beyond the
scope of this paper, but for a pedagogical treatment, see \cite{Pearson}.

\section{Final remarks and Conclusions}

The simplified method for determining the criticality threshold developed here employs very few approximations
and produces critical radii and critical masses within a few \% of values generally reported at high enrichment levels.
The primary approximations are with the energy independence of the neutron reaction, the nature of the transport cross sections, and the assumption that the neutron distribution is homogeneous in the sphere.  
Given the close agreement between this simplified calculation and the sophisticated precision calculation, these approximations appear to be valid and useful.   

An essential conclusion is that, within the present level of approximation, moderation effects on the relevant fast-neutron spectrum are modest: only a small adjustment of the effective $^{238}$U fission cross section is needed to bring the enrichment-dependent critical-mass estimates into reasonable agreement with more sophisticated MCNP calculations. This suggests that, in these fast assemblies, typical neutron lifetimes are short compared to the timescale for substantial spectral softening, and that in a strongly supercritical, rapidly diverging system the spectrum is continually replenished by freshly produced fission neutrons.

This formalism can be adapted to estimate neutron multiplication in general
threshold or subcritical assemblies.  Different geometries can be considered by determining the weighted mean free-path length for escape in the same manner as was done for a sphere. 
As an example of this, the application to a rod of $^{232}$Th subject to neutrino bombardment is presented in Appendix \ref{appth}.

Applying this analysis to more complex materials, such as uranium hydride, would require inclusion of the
energy change in scattering from hydrogen, in which the energy of the neutron is reduced by a factor of about four on average
(velocity reduced by a factor of 2) for fast neutrons. 

The energy dependence of the cross sections could be accounted for by following the time evolution of the neutron energy and breaking the energies into groups, as was done for more sophisticated calculations before MCNP; see ref. \cite{multigroup}.  However, the conclusion here is that this is a relatively small effect.

Finally, we compare Eq. (\ref{rc}) to the 1943-era Oppenheimer-Bethe critical radius formula presented as Eqs. (2) and (3) in \cite{chadwick1},
recast in the notation of Eq. (\ref{rc}):
$$
R_c=\frac{\pi}{\sqrt{3}} \sqrt{\frac{1}{\left[(v-1)  n^2 \sigma_{tr} \sigma_f\right]}} \times f.
$$
The factor
$f$ is a correction derived from a ``more accurate integral theory":
$$
f=\frac{1}{\left[1+0.9({v}-1) \sigma_f / \sigma_{tr}\right]}.
$$
Using the ENDF/B-VIII.0 values at 1.5 MeV, this equation leads to a pure $^{235}$U critical mass of 55 kg, compared to 40.8 kg from Eq. (\ref{rc}), and 46 kg using MCNP6 (\cite{chadwick1}, Table I, p. S32).   
For $^{239}$Pu, the result is 8.6 kg at 1.5 MeV (mass is reduced with increasing energy), compared to 9.92 kg using Eq. (\ref{rc}) at 1.9 MeV, while MCNP6 yields a critical mass of 10.2 kg (\cite{chadwick1}, Table 2, p. S34). 
Thus, our Eq. (\ref{rc}) appears to be more accurate; moreover, its derivation is more transparent and it has no intrinsic approximations beyond energy-independent cross sections and transport determined by a single effective isotropic scattering cross section, which are approximations common to both. 
We note that any criticality formula that has $(\nu-1)$ instead of $(\sigma_f\nu/\sigma_0 -1)$ has employed an approximation of low absorption.

Attempts to reproduce the Oppenheimer-Bethe equation by an expansion of the logarithm present in Eq. (\ref{rc}) 
were not met with success despite the similarities between the two formulae.

\section*{Acknowledgments}

The author acknowledges scientific and editorial commentary and criticism from Prof. B. Cameron Reed, who also independently checked the numerical results. Dr. Alon Goldring pointed out several typographical errors and other lapses and asked penetrating questions that led to the rewriting of a section that presents the effect of isotopic purity on the critical mass. 

Dr. Mark B. Chadwick, C.O.O. of the Los Alamos National Laboratory Weapons Directorate and a Laboratory Fellow, provided critical feedback and compared Eq. (\ref{rc}) with his own determinations of the critical masses of $^{235}$U and $^{239}$Pu 
using ENDF/B-VIII.0 cross sections with MCNP6, and pointed out that the average neutron energy in a critical assembly of $^{239}$ Pu is closer to 1.9 MeV. He also brought to my attention the declassified Oppenheimer-Bethe criticality relationship, presented as Eqs. (1) and (2) in \cite{chadwick1}, reproduced in the concluding section of this manuscript.

This work was supported by Yale University.

\section*{AI Use}
The final document was sanitized by Overleaf's AI editor (Writefull), which identified and repaired---subject to my approval---a few typographical errors, grammatical lapses, and run-on sentences. No content was generated by AI; it was used solely to rearrange the deck chairs on the Titanic.

\section*{Author Declarations}
I have no conflicts to disclose.

\newpage
\appendix

\section*{Supplemental Material}
These appendices contain supplemental material presenting derivations of needed mathematical and physical relationships that would be distracting if included in the narrative, but not directly essential to understanding the criticality derivation.

In addition, a new boundary condition is introduced for the diffusion equation approach that eliminates the transcendental equation that is usually encountered when solving the diffusion equation.  This simplifies a direct comparison with the elementary method obtained in the main narrative.

\section{Average squared distance to sphere surface}

\begin{figure}[h]

	\includegraphics[scale=0.6]{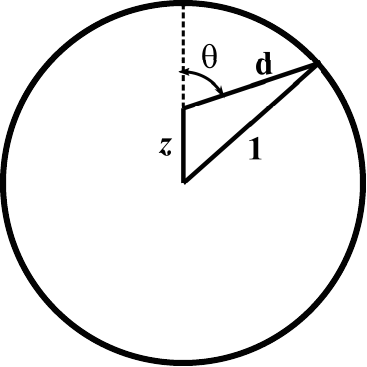}
 \caption{Geometrical picture for calculating the mean squared distance from any point inside a sphere to the surface.\label{circave1} }
\end{figure}

The average square distance from any point within a sphere to the surface of the sphere can be calculated as follows.  
Consider a point $z\leq 1$ on the $z$ axis of a three-dimensional Cartesian coordinate system within a unit sphere, 
with the geometry shown in Fig. \ref{circave1}.  The distance $\mathrm{d}$ from $z$ to the surface is independent of the azimuthal angle $\phi$ of the 
line from $z$ to the surface, but depends on the polar angle of the segment as
\[\mathrm{d}(\theta,z)=\sqrt{1-z^2\sin^2\theta}-z\cos\theta,\] 
which can be determined by the law of cosines and the quadratic formula.  Therefore, 
\begin{equation}
\mathrm{d}^2(\theta,z)=1-z^2\sin^2\theta +z^2\cos^2\theta-2z\cos\theta\sqrt{1-z^2\sin^2\theta}.
\end{equation}
 
The angle-averaged square distance is determined by  
$$\widebar{\mathrm{d}^2}(z)=\frac{1}{4\pi}\int d^2 (z,\theta) d\Omega = 1 - \frac{1}{3}z^2.$$ 

The odd-parity term in $\mathrm{d}^2(\theta,z)$ with $\cos\theta$ does not contribute to the integral over $\int_0^\pi \sin\theta\ d\theta$.  
Averaging over $z$ must include weighting by the volume of a spherical shell of radius $z$ and thickness $dz$
within a sphere of radius unity and volume $4\pi/3$: 
\begin{equation}
\widebar{\mathrm{d}^2}=3\int_0^1  z^2\left[1-\frac{1}{3}z^2\right] \, dz=\frac{4}{5}.
\end{equation}

This is for a unit radius sphere, and therefore in general
\begin{equation}
\widebar{\mathrm{d}^2}=\frac{4}{5}~ R^2.
\end{equation}

\section{Random walk correction}

The random walk displacement of Section IIC is an overestimate because larger displacements tend
to dominate the average of the squares. However, this can be taken into account and leads to a decrease in 
the displacement after a number of steps, which reduces the critical radius.  The displacement after $N_s$ steps is
\begin{equation}
\langle\ell\rangle=\left \langle \sqrt{N_s \ell_s^2}\right\rangle,
\end{equation}
which, noting that $\delta N_s=\sqrt{N_s}$, has standard deviation 
$$\sigma=\sqrt{N_s}~\ell_s.$$

Because $\ell_s$ is taken as a
constant, assume that $\ell$ is a proxy that carries the fluctuations in $N_s$ about its average, $N_s\pm\delta N_s$.
A weighted average can be formed as
\begin{eqnarray}
\langle\ell\rangle&=&
\sqrt{\sigma^2}\left\langle \sqrt {\frac{\ell^2}{\sigma^2}} \right\rangle\\ \nonumber
&=&\sqrt{N_s\ell_s^2}~ \left\langle\sqrt {\frac{\ell^2}{N_s\ell_s^2}}\right\rangle\\ \nonumber
&=&\sqrt{N_s\ell_s^2}~\left\langle\sqrt{X^2}\right\rangle,
\end{eqnarray}
where $X^2$ is a random variable with mean one and standard deviation one.  The average of this random variable is given by
the mean of the $\chi$ distribution (which is different from $\chi^2$; see \cite{johnson})
\begin{equation}
\left\langle \sqrt{X^2}\right\rangle=\sqrt{2}\,\frac{\Gamma\left((d+1)/2\right)}{\Gamma\left(d/2\right)}=\sqrt{\frac{2}{\pi}}=0.79\ {\rm for\ }d=1,
\end{equation}
where $d=1$ in the case of one random variable $X^2$.

This correction applies whenever an average length or distance is determined from the square root of a random-walk displacement.

Taking into account the $(4/5)R^2$ factor for the mean squared distance to the surface of a sphere, 
the overall correction factor for determining $R_c$ is
$$\epsilon=\sqrt{\frac{5}{4}}~\sqrt{\frac{2}{\pi}}=.8921.$$

The effect of sphere radius and random walk length has been extensively studied using a Monte Carlo analysis that validates this result.

\section{Estimation of the effect of forward scattering on the mean free path}

When neutrons of around 1 MeV energy elastically scatter from a high-$A$ nucleus, the scattering has a narrow forward-directed peak,
with a forward-to-backward ratio in the differential cross section of order 10 to 1.  Forward-directed scattering does not contribute significantly to the diffusive random-walk process. Historically, the elastic cross section was also given as a transport cross section,
 which includes an additional factor $\langle 1-\cos\theta\rangle$.
In modern nuclear tables it is rare to see the transport cross section listed with neutron scattering parameters. However, it can be estimated as follows.

The forward peak is due to the excitation of collective modes within the nucleus, of which there are many below 1 MeV. 
These excited modes can undergo an internal conversion that re-emits a neutron with lower energy (hence inelastic scattering) and roughly isotropically directed; they can also generate a directly produced coherent elastic wave in addition to the usual hard-core elastically scattered wave.

We can analyze the effect of the internally excited states by introducing a complex scattering length $i a_i$, in addition to the usual coherent elastic scattering length of the hard sphere $a_0$.  
As there are many levels that contribute to $a_i$, the total average elastic cross section is given by
$$\sigma_{tr}\propto |a_0+iMa_i|^2 \Delta\theta +a_0^2 + Ma_i^2,$$
where $M$ is the approximate number of levels that contribute and $\Delta \theta\approx 1/M$ is the width
of the forward peak where the interference between scattered waves due to internal processes add coherently. 
At larger angles, the contributions to the scattered wave add individually without the coherence enhancement.
For our purposes, these considerations lead to
\[|a_0+iMa_i|^2 \Delta\theta +a_0^2 + Ma_i^2=
\left[a_0^2+M^2a_i^2\right]\frac{1}{M} +a_0^2 + Ma_i^2\]
\[\approx Ma_i^2 + (a_0^2+Ma_i^2)=a_0^2+2Ma_i^2\propto\sigma_e,\]
while $$ Ma_i^2\propto \sigma_i.$$
For the average elastic cross section, internal excitations therefore contribute twice to the total cross section.

Accounting for the loss of effectiveness of the forward peak in the transport problem (multiplying the $Ma_i^2$ term $a_0^2$ by $\langle 1-\cos\theta\rangle\approx 1/2M^2$) 
reduces the total $\sigma_e$ by half of $\sigma_i$, which produces the relevant cross section for the neutron transport problem.
Therefore,
$$\sigma_{tr}\approx (\sigma_e-\frac{1}{2}\sigma_i)+\sigma_i\approx\sigma_e+\frac{1}{2}\sigma_i.$$

Although the excited internal-state matrix elements are not exactly the same for elastic scattering vs. the inelastic neutron
re-emission after internal conversion, they are likely of similar magnitude, so this would seem to be a reasonable approximation.  
In any case, the correction is small and enters as the square root.

This correction was found to be generally valid by a Monte Carlo study of the effects of the forward-directed component of the elastic scattering on a random walk, which of course is a purely mechanical study of the process and provides an estimate of $\langle 1- \cos\theta\rangle$.

A graph on p. S54 (Fig. 22) and Eq. (4) of \cite{chadwick1}  gives $\sigma_{tr}\approx 4.85$ b, compared to 4.52 b based on the analysis presented above. 
Also in \cite{chadwick1}, results from the PARTISN multigroup
transport code are provided. This code explicitly computes
a transport cross section and produces 4.83
b in a 618-group around 1.5 MeV and is consistent with the ENDF/B-VIII.0.
The successful implementation of these codes requires expertise that many non-professionals do not possess, in which case the simple estimate presented above is generally adequate. 

\section{Application to the diffusion equation}

The notion of the average path length squared used in the foregoing analysis can be replaced by considering the average path length as determined by the diffusion equation.

With the diffusion equation, the density of free neutrons depends on the radial position as $n(r)$, ignoring any angular dependence which is not expected for a steady state system.  In steady state,
$\nabla^2 n(r) =0$ and consider first the lowest order eigenmode of this equation, which will have the slowest time evolution for slight changes from equilibrium and therefore dominate the steady state solution. Setting a boundary condition that $n(r)=0$, the well-known solution is
\begin{equation}
	n(r)=n(0)\ \frac{\sin \pi r/R}{(\pi r /R)}.
\end{equation}
If there are $N$ neutrons in the system, we can normalize $n(r)$ to give $N$ as the integrated total,
\begin{equation}
	N=	4\pi n(0)\int_0^R \left[\frac{\sin \pi r/R}{(\pi r /R)}\right]r^2 dr= 4R^3
\end{equation}
and the density as a function of position is
\begin{equation}
	n(r)=\frac{ \pi N}{4 R^3}\left[\frac{\sin (\pi r/R)}{(\pi r /R)}\right].
\end{equation}

From the foregoing analysis, we know that neutrons are emitted at a rate $N/\tau_w$ for a system in equilibrium. However, the boundary condition above sets $n(R)=0$, which would seem to forbid neutron emission. Often, this is remedied by assuming a solution of the form $\sin\alpha r/\alpha r$, which leads to a transcendental equation in which the derivative at $R$ specifies the neutron loss rate.

The approach we take here is to add a background neutron gas of constant density $n_g$ that extends beyond the boundary, while maintaining the total $N$ within the sphere:
\begin{equation}
	n_g=\frac{\beta N}{(4/3)\pi R^3}.
\end{equation}
Since $\nabla^2 n_g=0$, we replace $n(r)$ by $(1-\beta)n(r)$, where $\beta$ is a small number. In some sense, this is equivalent to specifying the surface temperature in a thermal diffusion problem, and it is no more brutal than varying $\alpha$. Introducing $n_g$ enforces continuity of the density across the boundary. In free space, $n_g$ must then fall as $1/r^2$ for $r>R_c$, where $r$ is the distance from the center of the sphere.
    
If we consider an infinitesimally thin layer on the surface of the sphere, there is a flux of neutrons into this layer determined by the diffusion equation, but there is also a loss from this layer due to the background density $n_g$, which is given by kinetic theory.
In equilibrium, the radial diffusion flux matches the kinetic gas flow,
\begin{equation}
	-D \left[\nabla_r n(r)\right]\Big\vert_{r=R}= \frac{1}{4} n_g \widebar v
\end{equation}
where $D=\frac{1}{6} \widebar v \ell_s$ is the three-dimensional diffusion constant, and $n_g$ does not contribute to the diffusion flux. 
\begin{equation}
\beta\frac{ N}{(4/3)\pi R^3}\frac{\widebar v}{4}=(1-\beta)\frac{\widebar v \ell_s}{6}\frac{N}{4 R^3}\frac{\pi}{R}
\end{equation}
and which leads to 
\begin{equation}
\beta=\frac{4\pi^2}{18}\ \frac{\ell_s}{R}\ (1-\beta)
\end{equation}
and therefore
\begin{equation}
\beta=\frac{1}{1+(18/4\pi^2)(R/\ell_s)}.
\end{equation}
The rate of neutron emission is the flux times the area, and it is easier to use the kinetic flux, so the rate
\begin{equation}
\frac{\dot N}{N}=\frac{1}{\tau_w}=\frac{\widebar v}{\ell}=\beta\frac{ 1}{(4/3)\pi R^3}\ \frac{\widebar v}{4}\ 4\pi R^2
\end{equation}
and we arrive at
\begin{equation}
\frac{1}{\ell}=\frac{3}{4}\frac{\beta}{R}.
\end{equation}
Then the relation between $\ell$ and $R$ is
\begin{equation}
\ell=\frac{4}{3}R\left(1+\frac{18R}{4\pi^2\ell_s}\right).
\end{equation}
Letting $\ell'=\ell/\ell_s$ and $x=R/\ell_s$, the equation for $x$ is
\begin{equation}
-\ell'+\frac{4}{3}x+\frac{6}{\pi}x^2=0 
\end{equation}
which is not as simple as before, but readily solved.  

Of course we recall that $\ell$ is determined by the nuclear reaction properties, and the required $R$ to achieve that, when $\ell/\ell_s\sim 5$, is 
\begin{equation}
	R\approx 0.95\sqrt{\ell\ell_s}
\end{equation}
which is already similar to $\ell_c$ as derived regarding Eq. (\ref{rc}).  

There is one more correction, because $R$ is determined by a three-dimensional random walk through the diffusion equation, which determines the root mean square distance, whereas we need the mean of the square root. This can be determined as before from the $\chi$ distribution, and in general a $d$-dimensional random walk must be corrected by a factor\cite{S21}
\begin{equation}
	R_{msr}= R_{rms}\sqrt{\frac{2}{d}} \frac{\Gamma\left(\frac{d+1}{2}\right)}{\Gamma\left(\frac{d}{2}\right)}.
\end{equation}
Previously, we had $d=1$ because the random walk was in regard to the (average of) the square of the linear distance to the surface of the sphere (this turns the system into a 1-D problem).    For the diffusion equation, $d=3$, and we arrive at
\begin{eqnarray}
	R_c=\widebar{\sqrt{R^2}} &=&  \sqrt{\frac{2}{3}} \frac{\Gamma(2)}{\Gamma(3/2)}(0.95)\sqrt{\ell_c\ell_s}\\
    &=&(0.95)(0.921)\sqrt{\ell_c\ell_s}\\
    &=&0.875\sqrt{\ell_c\ell_s}=\epsilon'\sqrt{\ell_c\ell_s}.
\end{eqnarray}

Compared with Eq. (\ref{rc}), where $\epsilon=0.892$ we see a very satisfying agreement with $\epsilon'=0.875\approx\epsilon$.  Note that $\ell_c$ is completely determined by the nuclear properties, so only the mechanics of neutron transport are required to compare the two methods.

It might be argued that the diffusion approach avoids the weird numerology introduced in the foregoing random walk treatment. The equivalence of the results illustrates the general validity of the random-walk approach.

\section{Estimating the neutrino-induced fission neutron multiplication factor and escape probability in a rod of $^{232}$Th}\label{appth}

An important experiment seeks the induction of the fission of a heavy nucleus via a neutrino interaction. 
This is done by placing a large sample of $^{232}$Th in the form of a long rod near a reactor
and measuring the rate at which neutrons are produced by the sample due to the neutrino flux of the reactor.\cite{S11}

In this case, the calculation proceeds in a slightly different manner than for the critical mass determination. 
It is more profitable to view this as a time-dependent system with an initial event that causes a fission of a
$^{232}$Th nucleus, releasing a pulse of neutrons.  
The neutron-induced fission cross section of $^{232}$Th is relatively small, whereas the absorption cross section
(without inducing a fission) is approximately the same value.
In addition, there is a neutron energy threshold to induce fission of about 2 MeV, similar to that of $^{238}$U.\cite{S12} 

Parameters for $^{232}$Th given in \cite{S13} are  $\sigma_a= 0.096$ b, $\sigma_{tr}=4.83+2.69=6.18$ b,
$\sigma_f= 0.080$ b, $\nu =   2.4$, and $n =    3.04\times 10^{22}$ cm$^{-3}$.
Taking these together with $\langle R\rangle =10$ and $\sqrt{\langle R^2\rangle}=15 $ cm
leads to $\ell_s=5.3$ cm.

The number of scatterings for a neutron to random walk out of the rod is
 \begin{equation}
 N_s=\langle R^2\rangle/\ell_s^2\approx 225/5.3^2\approx 8,
 \end{equation}
 so the total distance traveled is
\begin{equation}
\ell\approx N_s\ell_s\approx 40\ {\rm cm}.
\end{equation}
The probability to escape from the rod (without absorption) is given by
\begin{equation}
P=e^{-n\sigma_a\ell}\approx e^{-0.12}\approx 0.89.
\end{equation}
implying $2.4\times 0.89\approx 2.1$ neutrons per induced fission that escape,
assuming $\sigma_f$ is not significant due to the energy loss per scattering.
There is a slight chance that the neutrons from the initial pulse can induce additional neutrons; however, the probability is low due to the 2 MeV fission threshold for $^{232}$Th.
The neutron energy loss per scattering is about 1-2\%, so after 10 scatterings the initial
neutron spectrum falls below the threshold.  
A rough estimate assuming a 10\% chance for secondary fissions brings the estimate to 1.3 neutrons per neutrino-induced fission.

Of course, a full Monte Carlo simulation is required for an accurate estimate of this dynamic system, 
but this back-of-the-envelope calculation provides a quick means for a sanity check when a program such as MCNP is employed.  

\clearpage

\section{Derivation of Eq. (5) by use of rate equations}

 \begin{figure}[H]
 	\includegraphics[width=0.9
 	\textwidth]{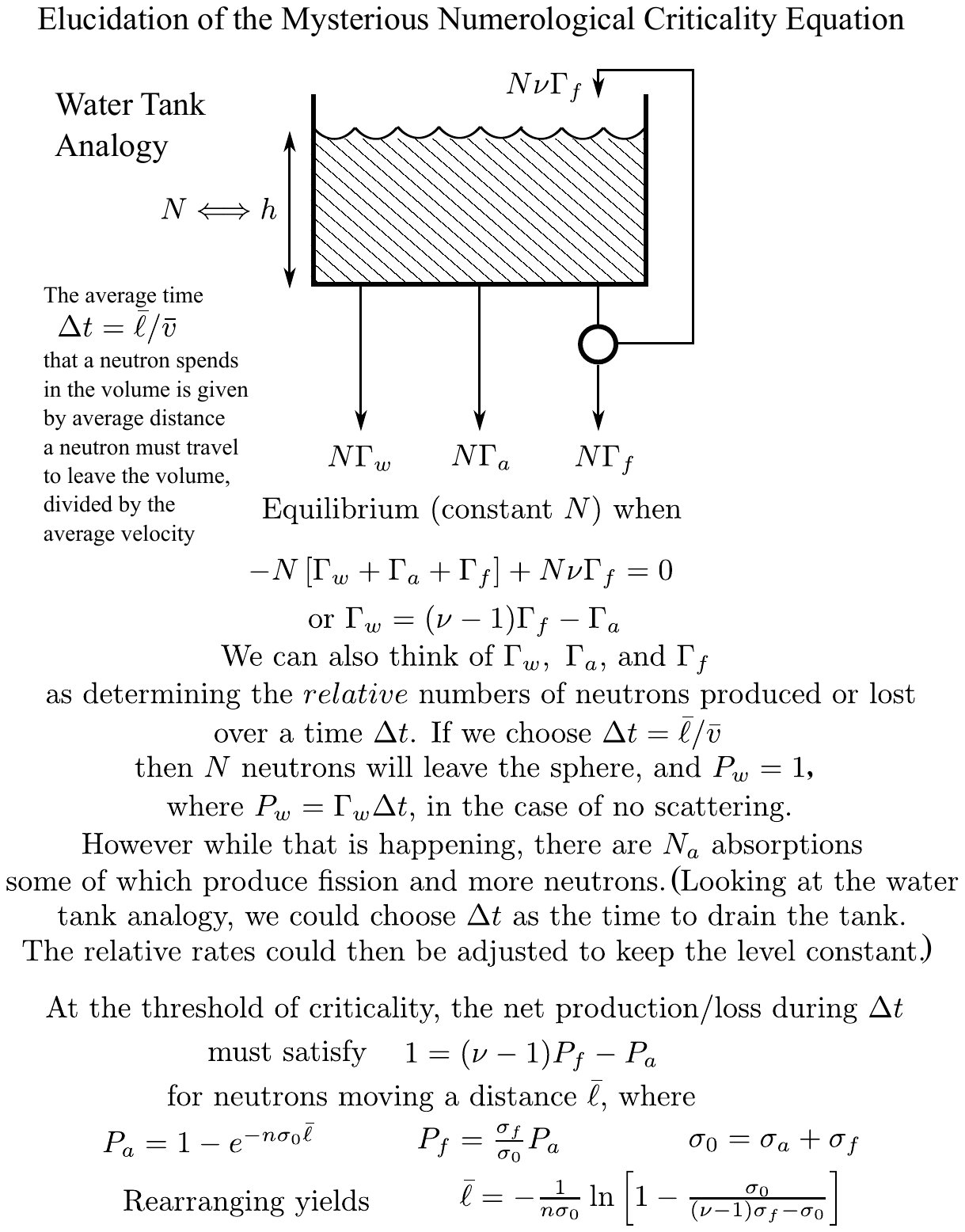}
 	
 \end{figure}
 \clearpage
\FloatBarrier

	\bibliographystyle{unsrt}

\vfill\eject

\end{document}